\documentclass[aps,prb,twocolumn,floatfix,groupedaddress,longbibliography]{revtex4-2}

\usepackage{amssymb}
\usepackage{amsmath}
\usepackage{bm}
\usepackage{graphicx}
\usepackage{dcolumn}
\usepackage[colorlinks,citecolor=red,linkcolor=blue,urlcolor=blue]{hyperref}

\usepackage{xcolor}
\usepackage{physics}
\usepackage{siunitx}

\allowdisplaybreaks 

\renewcommand\vec{\mathbf} 
\newcommand{\gvec}[1]{\bm{#1}} 
\newcommand{\hc}{\text{H.c.}}

\newcommand{\overbar}[1]{\mkern 1.5mu\overline{\mkern-1.5mu#1\mkern-1.5mu}\mkern 1.5mu}

\begin{document}

\title{Exact staggered dimer ground state and its stability in a two-dimensional magnet}
\author{Manas Ranjan Mahapatra}
\author{Rakesh Kumar}
\email[]{rkumar@curaj.ac.in}
\affiliation{School of Physical Sciences, Central University of Rajasthan, Ajmer 305817, India}

\date{\today}

\begin{abstract}
  Finding an exact solution for a realistic interacting quantum many-body problem is often challenging. There are only a few problems where an exact solution can be found, usually in a narrow parameter space. Here, we propose a spin-$1/2$ Heisenberg model on a square lattice with spatial anisotropy and bond depletion for the nearest-neighbor antiferromagnetic interactions but not for the next-nearest-neighbor interactions. This model has an \emph{exact} and \emph{unique} dimer ground state at $J_2/J_1=1/2$; a dimer state is a product state of spin-singlets on dimers (here, staggered nearest-neighbor bonds). We examine this model by employing the bond-operator mean-field theory and exact diagonalization. These analytical and numerical methods precisely affirm the correctness of the dimer ground state at the exact point ($J_2/J_1=1/2$). As one moves away from the exact point, the dimer order melts and vanishes when the spin gap becomes zero. The mean-field theory with harmonic approximation indicates that the dimer order persists for $-0.35\lesssim J_2/J_1\lesssim 1.35$. However, in non-harmonic approximation, the upper critical point lowers by $0.28$ to $1.07$, but the lower critical point remains intact. The exact diagonalization results suggest that the latter approximation fares better. The model reveals N\'eel order below the lower critical point and stripe magnetic order above the upper critical point. It has a topologically equivalent model on a honeycomb lattice where the nearest-neighbor interactions are still spatial anisotropic, but the bond depletion shifts into the isotropic next-neighbor interactions. Moreover, these models can also be generalized in the three dimensions.
\end{abstract}

\pacs{}

\maketitle

\section{Introduction} 
\label{sec:introduction}

Frustrated magnetism is a fascinating and vigorous field of research that attracts theoreticians, experimentalists, and material scientists equally due to emerging new understandings and challenges in matters with competing interactions~\cite{Anderson1987, Manousakis1991, Auerbach1994, Balents2010, Lacroix2011, Starykh2015, Diep2020, Broholm2020, Paschen2021}. Interacting spin systems with the valence-bond ordered ground states is one subclass of this field, which has been in focus for many years due to their novel physics about phases and their transitions. A straightforward prototype spin system with competing interactions is the Majumdar-Ghosh (MG) model~\cite{Majumdar1969a,*Majumdar1969b}. It is a spin-$1/2$ Heisenberg model on a one-dimensional chain lattice with nearest- and next-nearest-neighbor antiferromagnetic exchange interactions $J_1$ and $J_2$, respectively. Its ground state energy at  $J_2/J_1=1/2$ (MG point) is doubly degenerate, and the two \emph{nearly} orthogonal dimer states span the ground state manifold. As one increases the interaction ratio $J_2/J_1$, this model undergoes a quantum phase transition from a gapless quasi-long-range ordered state to a gapped dimer state at a quantum critical point, $J_2/J_1\sim 0.24$~\cite{Okamoto1992, Chitra1995}. Many quasi-one-dimensional materials show the signature of having the characteristic features of the MG model. The multiferroic compound CuCrO$_4$  finds its place close to the MG point~\cite{Law2011}, and the zigzag antiferromagnet (N$_2$H$_5$)CuCl$_3$ seems to be at the critical point~\cite{Maeshima2003}.

The existence of a true long-range order in two-dimensional spin systems at absolute zero temperature makes these systems more attractive than the one-dimensional systems. The spin-$1/2$ Heisenberg antiferromagnet with nearest-neighbor interaction ($J_1$) on a square lattice demonstrates the N\'eel magnetic order in its ground state. When it also contains an antiferromagnetic interaction ($J_2$) on the next-nearest-neighbor orthogonal bonds (or dimers) as specified in Ref.~\cite{Shastry1981}, then the resulting system, which is known as the Shastry-Sutherland (SS) model, yields an exact and unique dimer ground state for $J_1/J_2\leq1/2$~\cite{Shastry1981, Miyahara2003}. Moreover, rigorous analyses confirm that the SS dimer state remains an energetically favorable state up to  $J_1/J_2\sim 0.68$~\cite{Koga2000, Corboz2013}. As this exchange ratio increases beyond $0.76$, the antiferromagnetic N\'eel state emerges as the ground state. In the intermediate region, most studies confirm a plaquette spin-singlet state~\cite{Koga2000, Chung2001, Lauchli2002, Corboz2013, Lee2019, Yang2022, Xi2023}. The phase transition between the SS dimer and the plaquette states is of the first order. In contrast, a second-order transition with a deconfined quantum critical point is potentially possible between the plaquette and N\'eel phases~\cite{Lee2019}. Recently, a tensor network method-based probe found a spin supersolid phase in a narrow region just above the plaquette phase~\cite{Wang2023}. The structural and physical properties of SrCu$_2$(BO$_3$)$_2$ compound make it an excellent experimental realization of the SS model~\cite{Miyahara1999, Kageyama2000, Zayed2017, McClarty2017, Guo2020, Jimenez2021}.

There is another class of Hamiltonians in which frustration (as in MG and SS models) and spatial anisotropy play essential roles. One such example is the Nersesyan-Tsvelik (NT) model~\cite{Nersesyan2003}, which transforms to an isotropic $J_1$-$J_2$ Heisenberg antiferromagnet~\cite{Read1990} in the absence of anisotropy. The NT model consists of a set of identical spin chains that are arranged horizontally one after the other in two-dimensional space. Each spin chain carries exchange coupling $J_1^{\parallel}$ for any pair of neighboring spins of the chain and interacts with its nearest-neighboring spin chains through transverse and diagonal exchange couplings, denoted by $J_1^{\perp}$ and $J_2$, respectively. In this model, all couplings are antiferromagnetic and satisfy the condition $J_1^{\perp}, J_2\ll J_1^{\parallel}$. The line $J_1^\perp=2 J_2$ classically represents a first-order phase transition between the N\'eel and Stripe magnetic orderings. However, the analytical and numerical studies encapsulating the quantum nature of spins reveal a dimer phase between the N\'eel and Stripe orderings, and the dimer phase undergoes first-order transitions with both magnetically ordered phases~\cite{Sindzingre2004, Starykh2004}. The compound $(\text{NO})[\text{Cu}{({\text{NO}}_{3})}_{3}]$ appears a potential candidate for an experimental realization of the NT model~\cite{Volkova2010}.

Frustration and anisotropy are the primary sources for the emergence of dimer ground states in spin systems~\cite{Bose1991, Kumar2002, Takano2006, Mambrini2006, Gelle2008, Wenzel2008, Jiang2009, Kumar2009, Albuquerque2011, Kumar2017, Ghosh2022, Sushchyev2023}. These states can also arise from dimerization~\cite{Wenzel2009,  Danu2017} and  unfrustrated but competing interactions~\cite{Sandvik2007, Banerjee2011}. Dimer states are also in central focus in the evolving subject of deconfined quantum criticality driven by non-Landau phase transition~\cite{Senthil2004, Sandvik2007, Ganesh2013, Wang2016, Lee2019}. Motivated by all these works, we propose a spin-1/2 Heisenberg antiferromagnet with an exact dimer ground state and investigate it using bond-operator mean-field theory and numerical exact diagonalization.

We organize the remaining sections of this research paper as follows. The model and its ground state with supporting exact diagonalization data are placed in Sec.~\ref{sec:model}. After that, the bond-operator mean-field calculations are presented in Sec.~\ref{sec:bond_operator_mean_field_theory}. Subsequently, the mean-field and exact diagonalization results with analyses are provided in Sec.~\ref{sec:results_and_discussion}. Finally, we conclude this work in Sec.~\ref{sec:conclusions}.


\section{Model} 
\label{sec:model}

\begin{figure}[h]
  \centering
    \includegraphics[width=0.9\textwidth]{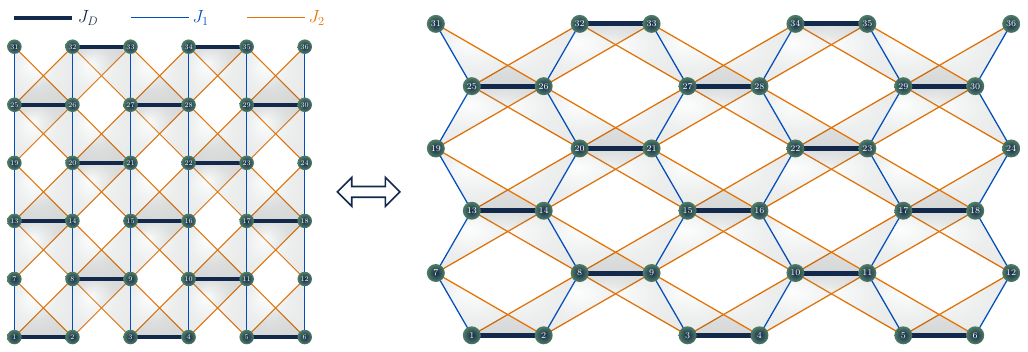}
  \caption{This picture represents a schematic representation of the model Hamiltonian~\eqref{eq:model_hamiltonian} on two topologically equivalent lattices. Here, the nearest-neighbor interactions have spatially anisotropy, but the second nearest-neighbor interactions are isotropic with the exchange coupling $J_2$. The horizontal nearest-neighbor bonds represented by thick lines are termed dimers. Each such dimer has interaction $J_D=2J_1 > 0$, where $J_1$ is the interaction strength on each of the remaining nearest-neighbor bonds. Filled circles with numbers denote the lattice sites where spin-$1/2$ degrees of freedom reside. Crossed shaded stripes can be visualized as the interpenetrating MG chains.}
  \label{fig:dimer_lattices}
\end{figure}

We consider the following spin-$1/2$ Hamiltonian, with periodic boundary conditions, for a SU(2) Heisenberg antiferromagnet on the two-dimensional lattice shown in Fig.~\ref{fig:dimer_lattices}:
\begin{equation}
  H = \sum_{\langle i,j\rangle}^{} J_{ij} \vec{S}_i \cdot \vec{S}_j + J_2 \sum_{\langle\langle i,j\rangle\rangle}^{} \vec{S}_i \cdot \vec{S}_j,
  \label{eq:model_hamiltonian}
\end{equation}
where $J_{ij}>0$ are on the nearest-neighbor bonds and have spatial anisotropy. A selected number of nearest-neighbor bonds, which we call dimers, has the coupling strength $J_D=2J_1$ (see thick blue bonds in Fig.~\ref{fig:dimer_lattices}). The remaining nearest-neighbor bonds have exchange interaction $J_1$. Furthermore, an isotropic spin-spin interaction exists between the spins connected by the next-nearest-neighbor bonds. This latter interaction can be ferromagnetic $(J_2<0)$ or antiferromagnetic ($J_2>0$). We define a dimensionless parameter $g=J_2/J_1$ and vary $J_2$ by setting $J_1=1$.

At the exact point ($g=1/2$) where an exact dimer ground state exists, the Hamiltonian~\eqref{eq:model_hamiltonian} can be expressed as
\begin{equation}
  H= \frac{3}{4}J_1 \sum_{\{(i,j,k)\}}^{} \mathcal{P}_{3/2}(i,j,k) - \frac{3}{2}J_1 N,
  \label{eq:exact_hamiltonian}
\end{equation}
where the summation runs over all possible triangles with vertices (or sites) $i, j$, and $k$ such that each triangle contains one dimer, one non-dimer nearest-neighbor bond, and one next-nearest-neighbor bond. If there are $N$ total dimers in the lattice with periodic boundary conditions, the total triangles will be $4N$. Equivalently, it means that each dimer contributes four triangles (see Fig.~\ref{fig:dimer_lattices}). The spin operators $\vec{S}_i, \vec{S}_j$, and $\vec{S}_k$ present at the vertices of a triangle define the projection operator $\mathcal{P}_{3/2}(i,j,k)$ as follows:
\begin{equation}
  \mathcal{P}_{3/2}(i,j,k)\equiv \frac{1}{3}(\vec{S}_i+\vec{S}_j+\vec{S}_k)^2 - \frac{1}{4}.
  \label{eq:projector}
\end{equation}
This operator projects a state of three spins localized at sites $i, j,$ and $k$ onto the $S=|\vec{S}_i+\vec{S}_j+\vec{S}_k|=3/2$ subspace. Three spin-$1/2$ degrees of freedom form two spin doublets and one spin quartet, which are the eigenstates of $\mathcal{P}_{3/2}$ with eigenvalues $0$ and $1$, respectively. This implies that the projection operator can be rewritten as $ \mathcal{P}_{3/2}=\sum_{M_z}^{} \op{S=3/2,M_z}$, and it annihilates a spin doublet state. When each triangle has a spin-doublet state, the Hamiltonian~\eqref{eq:exact_hamiltonian} gives the ground state energy, $E_{\text{GS}}=- \frac{3}{2}J_1 N$.

In general, there will be four spin-doublet states for a single triangle. This leads to ambiguity about the degeneracy and the type of doublet states for the ground state energy. To resolve these subtle issues, we first understand how the doublets form in a spin-triad. Coupling two spin-$1/2$ degrees of freedom generates a spin-singlet and three spin-triplet states. When these states are coupled with the states of the third spin, one gets four spin-doublet states and four spin-quartet states. The doublet states arising from the singlet state are ``separable'' in the sense that these can be expressed as a direct product of the spin-singlet and a state of the third spin. However, the remaining doublet states form entangled states between the spin-triplet states and two polarizations of the third spin.

If we start with an entangled spin-doublet state on a triangle, the remaining all $(4N-1)$ triangles will not be \emph{simultaneously} in the doublet states. On the other hand, if a dimer is in the spin-singlet state, all four triangles that share this dimer edge will be in the separable spin-doublet states. This results in the separable free one-spin states at the end sites of the other four neighboring dimers. Now, again, we can form spin-singlet states on the latter dimers. This leads to doublet states on the triangles containing these dimers and free one-spin states. We continue this process until all tringles are exhausted. The twofold degeneracy of the separable doublet on a triangle reduces to one as the free spins are constrained to form spin-singlet states on the dimers. Thus, at the exact point, the Hamiltonian~\eqref{eq:model_hamiltonian} has a unique ground state, a product of spin-singlet states on the dimers. This state can be mathematically written as
\begin{equation}
  \ket{\Psi} = \otimes_{(ij)\in\{\text{Dimers}\}}[i,j],
  \label{eq:exact_ground_state}
\end{equation}
where the direct product is over all dimers, $(ij)$ denotes a dimer with sites $i$ and $j$,  and $[i,j]\equiv\left(\ket{\uparrow_i \downarrow_j} - \ket{\downarrow_i \uparrow_j}\right)/\sqrt{2}$ is a spin-singlet state. A more formal and mathematical proof of the uniqueness of the ground state is rather rigorous and difficult, and a similar proof of our problem may be designed as found for the MG chain~\cite{Caspers1984}. We will not attempt it in this paper. Rather, we present exact-diagonalization data for the Hamiltonian~\eqref{eq:model_hamiltonian} on finite clusters below.

\onecolumngrid 

\twocolumngrid

\begin{figure}[h]
  \centering
    \includegraphics[width=.45\textwidth]{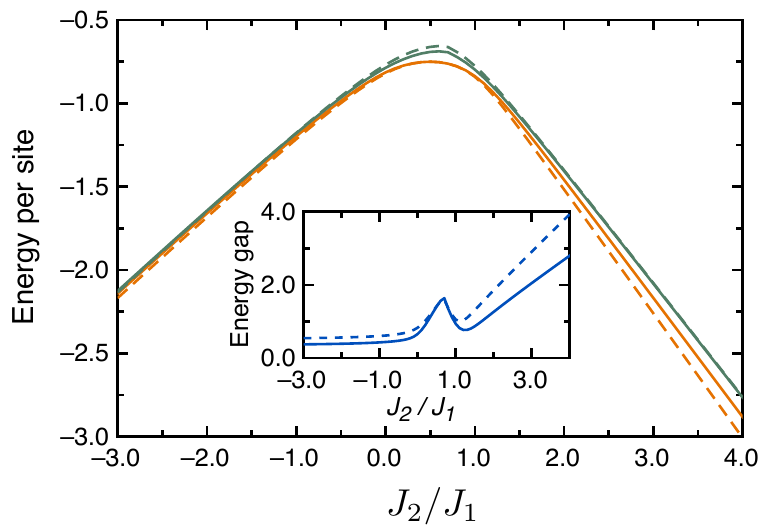}
  \caption{This picture displays energy per site for the ground and first excited states obtained from the exact diagonalization. The inset figure shows the gap between these two energies. The solid lines represent data for the $6 \times 4$ cluster, whereas dashed lines are for the $4 \times 4$ cluster.}
  \label{fig:ed_energies}
\end{figure}

\begin{figure}[h]
  \centering
    \includegraphics[width=.45\textwidth]{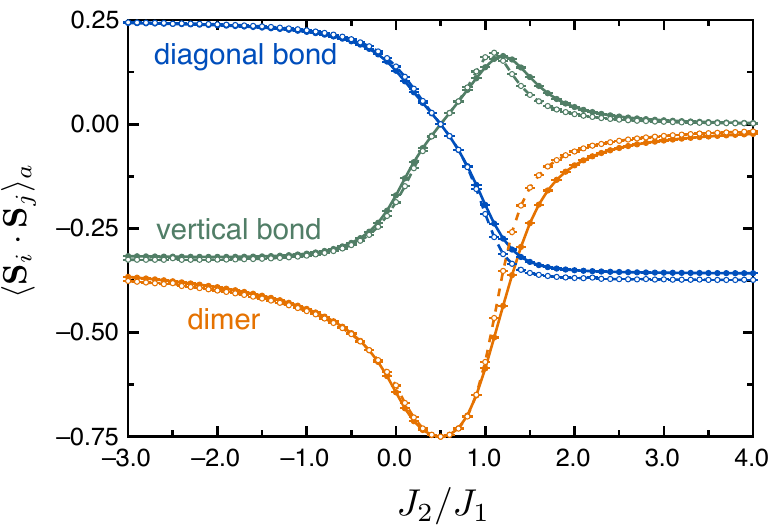}
  \caption{The averaged spin-spin correlations $\ev{\vec{S}_i \cdot \vec{S}_j}_{a}$ on dimers (or horizontal bonds), non-dimer nearest-neighbor bonds (or vertical bonds), and the next-nearest-neighbor bonds (or diagonal bonds) with error bars are shown here. Again, the solid lines belong to the $6 \times 4$ cluster, while the dashed lines belong to the $4 \times 4$ cluster.}
  \label{fig:ed_scorr_bond}
\end{figure}

We perform Lanczos-based diagonalization on square and rectangular lattices with even linear sizes in horizontal and vertical directions. An odd linear size along the vertical direction will break the periodic boundary conditions, and the staggered dimer state~\eqref{eq:exact_ground_state} is impossible with an odd linear size along the horizontal direction. The smallest $2 \times 2$ cluster has a self-overlapping problem, so we excluded it. We present the two lowest energies and the gap between them in Fig.~\ref{fig:ed_energies} for $4 \times 4$ and $6 \times 4$ clusters.
This data agrees perfectly with the theoretical prediction of a unique and exact dimer ground~\eqref{eq:exact_ground_state} at the exact point. Moreover, it also reflects that the ground energy remains non-degenerate as the system moves away from $J_2/J_1=0.5$. The energy gap data reveals asymmetry about the exact point. We also calculate the averaged spin-spin correlations on three types of bonds (see Fig.~\ref{fig:ed_scorr_bond}), which are horizontal bonds (or dimers), vertical bonds (or non-dimer nearest-neighbor bonds), and diagonal bonds (or next-nearest-neighbor bonds).
Spin-singlet formation (i.e., $\ev{\vec{S}_i \cdot \vec{S}_j}=-3/4$) on \emph{all} dimers and vanishing spin-spin correlation on other remaining bonds at $J_2/J_1=0.5$ leads to a product ground state. The magnitude of spin-spin correlation on dimers decreases on either side of the exact point. This behavior implies that a product ground state exists \emph{only} at the exact point. The spin-spin correlation on diagonal bonds saturates to its maximum positive value as one moves sufficiently left from $J_2/J_1=0.5$, which indicates ferromagnetic alignments along the diagonal bonds. On the other hand, as one increases $J_2$, all the correlations asymptotically saturate to values between the bounds $-3/4$ and $1/4$. Negligible error bars on the correlations reflect that all bonds of a type are equivalent and get the same spin-spin correlation value for all $J_2/J_1$.

\begin{figure}[h]
  \centering
    \includegraphics[width=.45\textwidth]{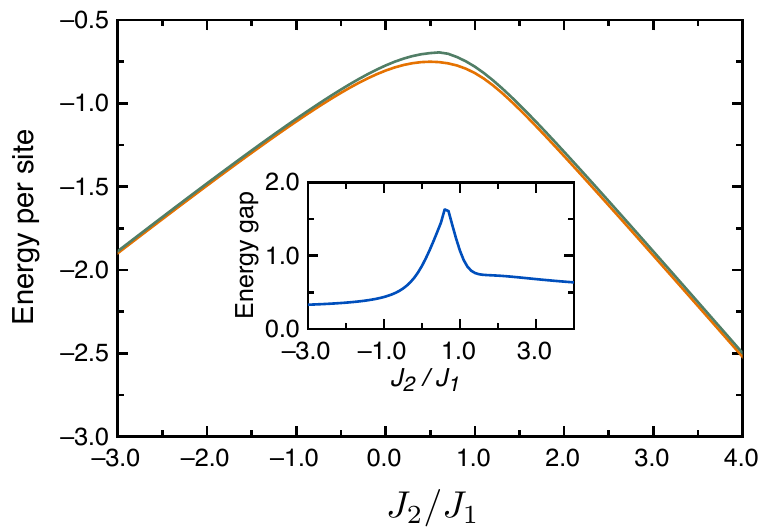}
  \caption{Two low-lying energies and the gap between them are given here from the exact diagonalization for the $4 \times 7$ cluster with open boundary conditions along the vertical direction.}
  \label{fig:ed_energiesOBC}
\end{figure}

\begin{figure}[h]
  \centering
    \includegraphics[width=.45\textwidth]{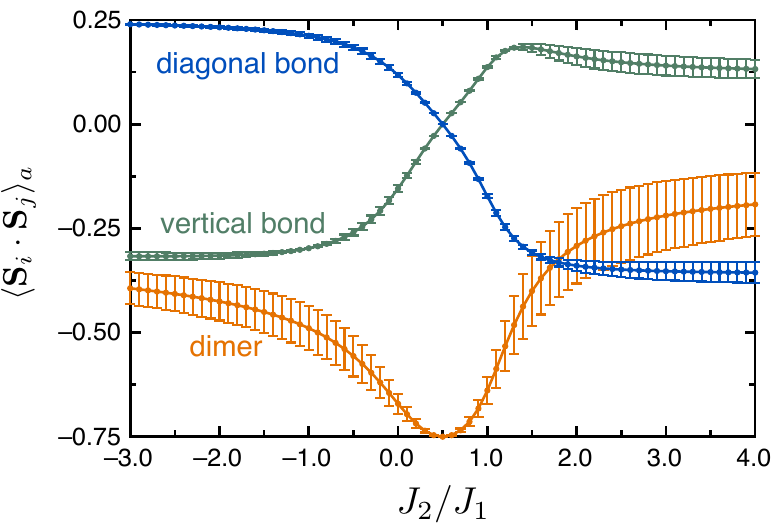}
  \caption{The averaged spin-spin correlation on the three types of bonds for the $4 \times 7$ cluster with open boundaries along the vertical direction is shown here. The error bar is negligible at the exact point but significantly increases on dimers as one goes farther away from $J_2/J_1=0.5$.}
  \label{fig:ed_scorr_bondOBC}
\end{figure}

The periodic boundary conditions along the vertical direction are not essential to get the product dimer ground state at the exact point. Furthermore, the linear size along this direction can be taken even as well as odd if one chooses open boundary conditions. However, even linear size and periodic boundary conditions along the horizontal direction are necessary for the staggered dimer ground state. In Fig.~\ref{fig:ed_energiesOBC}, we present two low-lying energies of the model~\eqref{eq:model_hamiltonian} for the $4 \times 7$ cluster with cylindrical boundary conditions.
At $J_2/J_1=0.5$, the ground state energy per dimer is $-3/4$ (in units of $J_D$), same as for the staggered dimer ground state~\eqref{eq:exact_ground_state}. The averaged spin-spin correlation data presented in Fig.~\ref{fig:ed_scorr_bondOBC} confirms the dimer ground state at the exact point. 
All bonds of a type make the same contribution at the exact point in cylindrical boundary conditions, too. However, as one moves away from $J_2/J_1$, distinct contributions from the dimers become prominent, and noticeable changes emerge in vertical and diagonal bonds.

\begin{figure}[h]
  \centering
    \includegraphics[width=.45\textwidth]{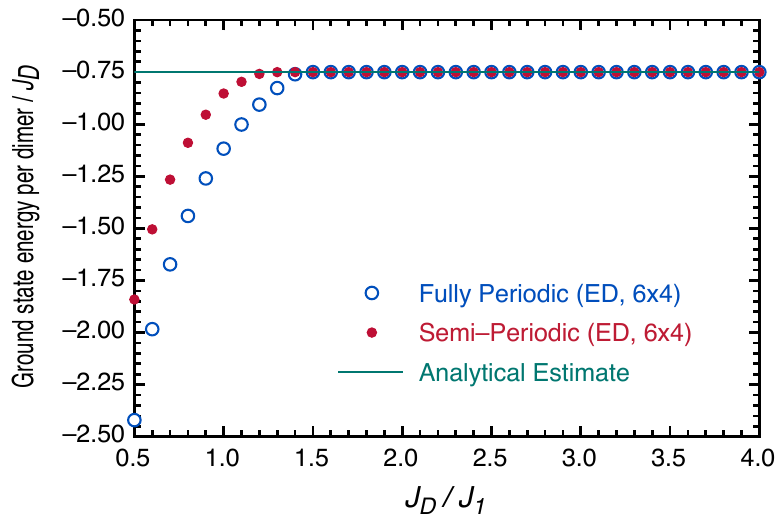}
  \caption{This figure displays the ground state energy per dimer (in units of $J_D$) calculated from the exact diagonalization and analytical calculations; here, we keep the exchange constant $J_1$ to unity. The diagonalization data is produced for the $6 \times 4$ lattice with fully- and semi-periodic boundary conditions; here, semi-periodic means boundaries along the vertical direction are open.}
  \label{fig:eng_est_exact}
\end{figure}

It would be interesting to compare the model~\eqref{eq:model_hamiltonian} with the MG and SS models. All these models involve only nearest-neighbor and next-nearest-neighbor exchange interactions and have dimer ground states at specific points or in a range in the coupling space. Except for the MG model, the dimer ground states are unique. Our and SS models have a spatial anisotropy in the nearest-neighbor and next-nearest-neighbor interactions, respectively. However, the MG model is isotropic in both interactions. Structurally, the SS and our models correspond to the bond-depleted $J_1$-$J_2$ Heisenberg models on a square lattice. One wipes out $3/4$th next-nearest-neighbor bonds in the SS case, while we eradicate $1/4$th nearest-neighbor bonds in our spatially anisotropic (or dimerized) model. Moreover, our model also shows the staggered dimer ground state for any $J_D \geq 2J_1$ if one sets the ratio $J_2/J_1$ equal to $1/2$, as the SS ground state in the SS model exists in a broad range. This can be understood as follows. At $J_2/J_1=1/2$, using Eq.~\eqref{eq:projector}, the Hamiltonian~\eqref{eq:model_hamiltonian} can be written as
\begin{equation}
  H=E_c +\frac{J_1}{2} \sum_{\{(i,j,k)\}}^{} \left[\kappa\,\vec{S}_i \cdot \vec{S}_j  + \frac{3}{2} \, \mathcal{P}_{3/2}(i,j,k)\right],
  \label{eq:hamil_jdtune}
\end{equation}
where $E_c=-(3/2)J_1N$, $\kappa=\left(J_D/2J_1-1\right)$, and the indices $i,j$ are endpoints of a dimer. The product state~\eqref{eq:exact_ground_state} is an eigenstate of the Hamiltonian~\eqref{eq:hamil_jdtune} with the eigenenergy $-(3/4) J_D N$, independent of the coupling $J_1$. As we know the slack inequalities $-3/4\leq \ev{\vec{S}_i \cdot \vec{S}_j}\leq 1/4$ and $0\leq\ev{ \mathcal{P}_{3/2}(i,j,k)}\leq 1$ of expectation values hold for any arbitrary state that is a linear combination of the energy eigenstates, the staggered dimer state will be the ground state for any $\kappa\geq 0$ (see Fig.~\ref{fig:eng_est_exact}).
It turns out that the dimer ground state also persists for a range with $\kappa<0$, similar to the SS model~\cite{Miyahara2003}. We arrive at the same conclusions of the Hamiltonian with fully-periodic boundary conditions even if the boundaries along the vertical directions are kept open, except that the range of $\kappa$ for the dimer state being a ground state is slightly wider in the semi-periodic case.


\section{Bond-operator mean-field theory} 
\label{sec:bond_operator_mean_field_theory}

This theory and its various variants are appropriate for the investigation of quantum spin systems in which the ground state is a product of spin singlets~\cite{Sachdev1990, Zhitomirsky1996, Kotov1998, Kumar2010, Syromyatnikov2018}. For spin-$1/2$ systems, a spin singlet forms when an even number of spins couple together. The present work considers only dimer spin singlets as our Hamiltonian has an exact staggered dimer ground state, and our objective is to examine the stability of this state in the coupling space.

Using the bond-operator formalism~\cite{Sachdev1990}, we transform a spin Hamiltonian from spin to bosonic Fock space representation. In the bosonic description, the creation operators $s^\dagger$, $t_x^\dagger$, $t_y^\dagger$, and $t_z^\dagger$ create boson particles of type $s$, $t_x$, $t_y$, and $t_z$, respectively, when applied on the vacuum $\ket{0}$ (a no particle state). To write Hamiltonian in terms of bosonic creation and annihilation operators, we define a \emph{one-to-one} mapping between the bosonic particle states and the spin states of a dimer as follows:
\begin{subequations}
  \label{eq:mapping}
  \begin{align}
    s^\dagger\ket{0}:=\ket{s}&\equiv \frac{1}{\sqrt{2}}(\ket{\uparrow \downarrow}-\ket{\downarrow \uparrow}),\label{eq:sdef}\\
     t_x^\dagger\ket{0}:= \ket{t_x}&\equiv\frac{-1}{\sqrt{2}}(\ket{ \uparrow \uparrow}-\ket{\downarrow \downarrow}),\label{eq:txdef}\\
    t_y^\dagger\ket{0}:=  \ket{t_y}&\equiv \frac{i}{\sqrt{2}}(\ket{ \uparrow \uparrow}+\ket{\downarrow \downarrow}),\label{eq:tydef}\\
     t_z^\dagger\ket{0}:=\ket{t_z}&\equiv\frac{1}{\sqrt{2}}(\ket{\uparrow \downarrow}+\ket{\downarrow \uparrow}),\label{eq:tzdef}
  \end{align}
\end{subequations}
where the states before the mapping symbol $:=$ are of different types of bosonic particles, and the remaining states are of coupled dimer spins.

The singlet and triplet states listed above in Eq.~\eqref{eq:mapping} form an orthonormal set of simultaneous eigenstates of $\vec{S}_1 \cdot \vec{S}_2$ and spin-inversion operator ($\pi_s$), but the states~\eqref{eq:txdef} and~\eqref{eq:tydef} are not the eigenstates of the $z$-component of total spin operator ($S_z=S_{1z}+S_{2z}$). Here, the subscripts $1$ and $2$ label the ends of a dimer. Other orthonormal sets are also possible, as the degeneracy in triplet states gives rise to arbitrariness in choosing the dimer spin eigenstates. Therefore, the mapping~\eqref{eq:mapping} is \emph{not} unique (for example, the Ref.~\cite{Syromyatnikov2018} defines a different mapping).

Using second quantization, the one-body operator $\mathcal{O}=S_{m \alpha} (m=1,2; \alpha=x,y,z)$ in the Fock space reduces to the following form:
\begin{equation}
  \mathcal{O} = \sum_{\mu \nu}^{} \mel{\mu}{\mathcal{O}}{\nu} \mu^\dagger \nu,
\end{equation}
where $\mu,\nu\in\{s,t_x,t_y,t_z\}$. After finding out the matrix elements $\mel{\mu}{\mathcal{O}}{\nu}$, the spin operators $S_{1 \alpha}$ and $S_{2 \alpha}$ in the Fock space can be written as
\begin{align}
  S_{m \alpha} &=-\left[\frac{(-1)^m}{2}(s^\dagger t_{\alpha}+t_{\alpha}^\dagger s)+\frac{i}{2} \epsilon_{\alpha \beta \gamma}t_{\beta}^\dagger t_{\gamma}\right],
  \label{eq:spin2boson}
\end{align}
where $\alpha,\beta,\gamma\in\{x,y,z\}$, and $\epsilon_{\alpha \beta \gamma}$ is a totally antisymmetric (Levi-Civita) tensor. Here, we do summations over the repeated Greek indices. The dimensionality of the Fock space of bosonic operators $s,t_{\alpha}$ is, in principle, infinite. However, the Hilbert space of a spin dimer is four-dimensional and can be spanned by $\ket{s}$, $\ket{t_x}$, $\ket{t_y}$, and $\ket{t_z}$). Therefore, the four Fock states $\mu^\dagger\ket{0}$ only span the physical subspace. To wipe out the remaining unphysical subspace, one imposes the following hard-core constraint on each dimer:
\begin{equation}
  \label{eq:constraint}
  s^\dagger s+ t_{\alpha}^\dagger t_{\alpha}=1.
\end{equation}
Using Eqs.~\eqref{eq:spin2boson} and~\eqref{eq:constraint}, the intra-dimer and inter-dimer spin-spin interactions of the form $\vec{S}_i \cdot \vec{S}_j$ can be written as
\begin{itemize}
  \item for intra-dimer interactions ($ \vec{r}=\vec{r}^\prime$):
  \begin{equation}
    \label{eq:sisj_intra}
    S_{1 \alpha} (\vec{r}) S_{2\alpha}(\vec{r}^\prime)= -\frac{3}{4}s^\dagger(\vec{r})s (\vec{r})+ \frac{1}{4}t_{\alpha}^\dagger(\vec{r})t_{\alpha}(\vec{r}),
  \end{equation}
  \item for inter-dimer interactions ($\vec{r}\neq\vec{r}^\prime; m,m^\prime=1,2$):
  \begin{equation}
    \label{eq:sisj_inter}
    S_{m \alpha} (\vec{r}) S_{m^\prime \alpha}(\vec{r}^\prime)=\frac{1}{4} \sum_{p=2}^{4} T_{mm^\prime}^{(p)}(\vec{r},\vec{r}^\prime),
  \end{equation}
\end{itemize}
where $\vec{r}$ and $\vec{r}^\prime$ are the position vectors of dimers, and
\begin{align}
  T_{mm^\prime}^{(2)}(\vec{r},\vec{r}^\prime) &= (-1)^{m+m^\prime}\left[ s(\vec{r})s^\dagger(\vec{r}^\prime)t_{\alpha}^\dagger(\vec{r})t_{ \alpha}(\vec{r}^\prime)\right.\nonumber \\
  &\left.+s(\vec{r})s(\vec{r}^\prime)t_{\alpha}^\dagger(\vec{r})t_{ \alpha}^\dagger(\vec{r}^\prime)+\hc\right],\label{eq:t2expr}\\
  T_{mm^\prime}^{(3)}(\vec{r},\vec{r}^\prime) &= \epsilon_{\alpha \beta \gamma}\left[(-1)^m+(-1)^{m^\prime} P_{\vec{r}\vec{r}^\prime}\right]\nonumber\\
  &\times\left[i s(\vec{r})t_{\alpha}^\dagger(\vec{r})t_{\beta}^\dagger(\vec{r}^\prime)t_{ \gamma}(\vec{r}^\prime)+\hc\right],\label{eq:t3expr}\\
   T_{mm^\prime}^{(4)}(\vec{r},\vec{r}^\prime) &=t_{\alpha}^\dagger(\vec{r})t_{ \beta}(\vec{r})\left(1-P_{\alpha\beta}\right)t_{\beta}^\dagger(\vec{r}^\prime)t_{ \alpha}(\vec{r}^\prime).\label{eq:t4expr}
\end{align}
We used an exchange operator $P_{ij}$ in the above two expressions, which interchanges the indices $i$ and $j$. One notices here that the operators $ T_{mm^\prime}^{(2)}(\vec{r},\vec{r}^\prime)$ and $ T_{mm^\prime}^{(3)}(\vec{r},\vec{r}^\prime)$ are symmetric and anti-symmetric, respectively, under the exchange of $m$ and $m^\prime$ indices. Moreover, the operator $T_{mm^\prime}^{(4)}(\vec{r},\vec{r}^\prime)$ does not depend on $m$ and $m^\prime$, i.e., $T_{mm^\prime}^{(4)}(\vec{r},\vec{r}^\prime)=T^{(4)}(\vec{r},\vec{r}^\prime)$. Additionally, from Eqs.~\eqref{eq:t2expr} and~\eqref{eq:t3expr}, we get the following useful properties:
\begin{align}
  T_{11}^{(2)}(\vec{r},\vec{r}^\prime) &= T_{22}^{(2)}(\vec{r},\vec{r}^\prime)=-T_{12}^{(2)}(\vec{r},\vec{r}^\prime),\\
  T_{11}^{(3)}(\vec{r},\vec{r}^\prime) &= -T_{22}^{(3)}(\vec{r},\vec{r}^\prime).
\end{align}

\subsection{In the staggered dimer phase} 
\label{sub:in_staggered_dimer_phase}
Recasting the model Hamiltonian~\eqref{eq:model_hamiltonian} for the staggered dimer state~\eqref{eq:exact_ground_state}, on the bond-depleted square lattice (Fig.~\ref{fig:dimer_lattices}), as
\begin{align}
  \label{eq:hamil_dimer}
  H&=J_{D} \sum_{\vec{r}\in\mathcal{D}}^{}\vec{S}_1(\vec{r}) \cdot \vec{S}_2(\vec{r}) + H_1 + H_2,
\end{align} 
where $\mathcal{D}$ is a set of position vectors of the Bravais lattice (two-dimensional oblique lattice) corresponding to the lattice shown in Fig.~\ref{fig:dimer_lattices}. In the ``lattice with a basis'' language, the two dimer sites can be imagined as a basis such that the left end of the dimer coincides with the Bravais lattice vector $\vec{r}$. The $H_1$ and $H_2$ in Eq.~\eqref{eq:hamil_dimer} have the following forms:
\begin{align}
  H_1&=\frac{J_1}{2}\sum_{\vec{r}\in\mathcal{D}}^{} \sum_{\vec{d}_1=\pm a\hat{y}}^{}\sum_{m=1,2}^{} \vec{S}_m(\vec{r})\cdot\vec{S}_{\overbar{m}}(\vec{r}_m+\vec{d}_1),\\
  H_2 &=\frac{J_2}{2}\sum_{\vec{r}\in\mathcal{D}}^{} \sum_{\vec{d}_2=\pm a\hat{x}\pm a\hat{y}}^{}\sum_{m=1,2}^{} \vec{S}_m(\vec{r})\cdot\vec{S}_m(\vec{r}+\vec{d}_2),
\end{align}
where $\overbar{m}\equiv3-m$, and $\vec{r}_m\equiv\vec{r}+(-1)^m a\hat{x}$ with the lattice constant $a$ of the square lattice (Fig.~\ref{fig:dimer_lattices}). Using Eqs~\eqref{eq:sisj_intra},~\eqref{eq:sisj_inter}, and the properties of $T^{(p)}$ operators, the Hamiltonian~\eqref{eq:hamil_dimer} can be written as
\begin{align}
  \label{eq:hamil_boperators}
  H&=J_D \sum_{\vec{r}}^{} \left[-\frac{3}{4}s^\dagger(\vec{r})s(\vec{r}) + \frac{1}{4} t_{\alpha}^\dagger(\vec{r})t_{\alpha}(\vec{r})\right] \nonumber\\
  &+\frac{1}{4}\sum_{\vec{r},\vec{d}_2}^{} \left[J_{-}T_{11}^{(2)}(\vec{r},\vec{r}+\vec{d}_2)+J_{+}T^{(4)}(\vec{r},\vec{r}+\vec{d}_2)\right] \nonumber \\
  &+\frac{J_1}{8} \sum_{\vec{r},\vec{d}_1}^{} \sum_{m=1,2}^{} T_{m\overbar{m}}^{(3)}(\vec{r},\vec{r}_m+\vec{d}_1)\nonumber \\
  &- \sum_{\vec{r}}^{}\mu_{\vec{r}} \left[s^\dagger(\vec{r})s(\vec{r}) + t_{\alpha}^\dagger(\vec{r})t_{\alpha}(\vec{r})-1\right],
\end{align}
where $J_{\pm}=J_2\pm J_1/2$. In the above equation, constraint~\eqref{eq:constraint} is incorporated as in Lagrange’s multiplier method to ensure that dimer states are always from the physical subspace.

We have dimer translation symmetry in $H$ with translations $\vec{r}=n_{+}\gvec{\tau}_{+}+n_{-}\gvec{\tau}_{-}$, where $n_{\pm}\in\mathbb{Z}$ and the primitive vectors $\gvec{\tau}_{\pm}=a\hat{x}\pm a \hat{y}$. So, we can replace local chemical potential $\mu_{\vec{r}}$ by a global $\mu$. Moreover, at the exact point, we get a set of perfectly spin-singlet dimers that form the ground state~\eqref{eq:exact_ground_state}. The terms $-\frac{3}{4}J_D s^\dagger(\vec{r})s(\vec{r})$ in Eq.~\eqref{eq:hamil_boperators} clearly confirm this. Therefore, we assume that a single singlet boson condenses on each dimer, that is, $\ev{s}=\ev{s^\dagger}=\bar{s}$. As one moves away from the exact point, the condensate amplitude decreases. In harmonic approximation, we ignore the triplet-triplet interactions (equivalently, neglecting the $T^{(3)}$ and $T^{(4)}$ terms). We include these higher-order terms using quadratic mean-field decoupling for calculation beyond the harmonic approximation. In this case, the term $T^{(3)}$ vanishes after decoupling due to the presence of the antisymmetric Levi-Civita tensor, while the Hartree-Fock decoupling of a $T^{(4)}$ term gives
\begin{align}
    T^{(4)}(\vec{r},\vec{r}+\vec{d}_2)&\approx (Q^2 - P^2) + P \left[t_{\alpha}^\dagger(\vec{r})t_{\alpha}(\vec{r}+\vec{d}_2)+\hc\right]\nonumber\\
    &-Q\left[t_{\alpha}^\dagger(\vec{r})t_{\alpha}^\dagger(\vec{r}+\vec{d}_2)+\hc\right],
  \end{align}
  where $ P\equiv \ev{t_{\alpha}^\dagger(\vec{r})t_{\alpha}(\vec{r}+\vec{d}_2)}$ and $Q\equiv \ev{t_{\alpha}^\dagger(\vec{r})t_{\alpha}^\dagger(\vec{r}+\vec{d}_2)}$. Incorporating these approximations, we obtain the following mean-field Hamiltonian
\begin{align}
  H_{mf}&=\mathcal{C}+\mu_r \sum_{\vec{r}}^{}t_{\alpha}^\dagger(\vec{r})t_{\alpha}(\vec{r})\nonumber\\ &+\frac{1}{4}\sum_{\vec{r},\vec{d}_2}^{}\left\{\mathcal{A}\left[t_{\alpha}^\dagger(\vec{r})t_{\alpha}(\vec{r}+\vec{d}_2)+\hc\right]\right.\nonumber\\ &\left.+\mathcal{B}\left[t_{\alpha}^\dagger(\vec{r})t_{\alpha}^\dagger(\vec{r}+\vec{d}_2)+\hc\right]\right\},
\end{align}
where $\mu_r=\frac{J_D}{4}-\mu$, $\mathcal{A}=J_{-}\bar{s}^2+J_{+}P$, $\mathcal{B}= J_{-}\bar{s}^2-J_{+}Q$, and $\mathcal{C}/N= -\frac{3}{4}J_D\bar{s}^2+\mu(1-\bar{s}^2)+ J_{+}\left(Q^2-P^2\right)$, with the total number of dimers $N$.

Utilizing the dimer translation symmetry, we define the Fourier transform: $ t_{\alpha}(\vec{r})=\frac{1}{\sqrt{N}}\sum_{\vec{k}}^{} e^{i\vec{k}\cdot\vec{r}}t_{\vec{k}\alpha}$, where the summation goes over all allowed $N$ Bloch wave vectors $\vec{k}$ inside the first Brillouin zone (here, it is a magnetic zone). The mean-field Hamiltonian in $k$-space takes the form:
\begin{align}
  \label{eq:hamil_kspace} H_{mf}&=E_0+\frac{1}{2}\sum_{\vec{k}}^{}\left[\Lambda_\vec{k}\left(t_{\vec{k}\alpha}^\dagger t_{\vec{k}\alpha}+t_{-\vec{k}\alpha}t_{-\vec{k}\alpha}^\dagger\right)\right.\nonumber\\
  &\left.+\Delta_\vec{k}\left(t_{\vec{k}\alpha}^\dagger t_{-\vec{k}\alpha}^\dagger+\hc\right)\right],
\end{align}
where $\Lambda_{\vec{k}}=\mu_r+2\mathcal{A}\gamma_{\vec{k}}$, $\Delta_{\vec{k}}=2\mathcal{B} \gamma_\vec{k}$, and $E_0=\mathcal{C}-\frac{3}{2} \sum_{\vec{k}}^{}  \Lambda_{\vec{k}}$, with a geometric structure factor $ \gamma_{\vec{k}}=\cos(k_x a)\cos(k_y a)$. 

For diagonalizing the $H_{mf}$, we define the Bogoliubov
transformation: $t_{\vec{k}\alpha}=u_{\vec{k}}\gamma_{\vec{k}\alpha}+v_{\vec{k}}\gamma_{-\vec{k}\alpha}^\dagger$. This transformation is canonical if $u_{\vec{k}}$ and $v_{\vec{k}}$ satisfy $u_{\vec{k}}^2-v_{\vec{k}}^2=1$.  One also observes that the $u_{\vec{k}}=u_{-\vec{k}}$ and $v_{\vec{k}}=v_{-\vec{k}}$ hold as we have dimer translation symmetry. After transforming the Hamiltonian~\eqref{eq:hamil_kspace} from $t$-bosons to $\gamma$-bosons using the Bogoliubov transformation, we get
\begin{equation}
  \label{eq:hamil_diag}
  H_{diag}=E_0 +\frac{3}{2}\sum_{\vec{k}}^{}\omega_{\vec{k}}   +\sum_{\vec{k}}^{}\omega_{\vec{k}}\gamma_{\vec{k}\alpha}^\dagger \gamma_{\vec{k}\alpha}
\end{equation}
provided the condition $2u_{\vec{k}}v_{\vec{k}}/(u_{\vec{k}}^2+v_{\vec{k}}^2) =-\Delta_{\vec{k}}/\Lambda_{\vec{k}}$ holds for all $\vec{k}$ points. The quasi $\gamma$-bosons  (often called as ``triplons'') disperse in the singlet background with the energy $\omega_{\vec{k}}=\sqrt{\Lambda_{\vec{k}}^2-\Delta_{\vec{k}}^2}$.

We extract the ground state properties from~\eqref{eq:hamil_diag} by finding the self-consistent equations that come after taking the partial derivatives of the ground state energy $ E_g= E_0+\frac{3}{2}\sum_{\vec{k}}^{}\omega_{\vec{k}}$ with respect to mean-field parameters $\mu$, $\bar{s}$, $P$, and $Q$. After simplification, we get the following self-consistent equations:
\begin{align}
  \bar{s}^2 &= \frac{5}{2} - \frac{3}{2N} \sum_{\vec{k}}^{} \frac{\Lambda_{\vec{k}}}{\omega_{\vec{k}}},\label{eq:sce_ss}\\
  \mu &=-\frac{3}{4}J_D+\frac{3}{N} J_{-}\sum_{\vec{k}}^{} \frac{\gamma_{\vec{k}} (\Lambda_{\vec{k}}-\Delta_{\vec{k}})}{\omega_{\vec{k}}},\label{eq:sce_mu}\\
  P &= \frac{3}{2N}  \sum_{\vec{k}}^{} \frac{ \gamma_{\vec{k}}\Lambda_{\vec{k}}}{\omega_{\vec{k}}},\label{eq:sce_p}\\
  Q &= -\frac{3}{2N}  \sum_{\vec{k}}^{} \frac{ \gamma_{\vec{k}}\Delta_{\vec{k}}}{\omega_{\vec{k}}}.\label{eq:sce_q}
\end{align}
We find the solution of these equations in the dimer phase at all suitable $J_2/J_1$ ratios by an iterative method as in Ref.~\cite{Kumar2008} or by optimization methods like the Levenberg–Marquardt algorithm~\cite{Levenberg1944, Marquardt1963}.


\subsection{In the ordered phases} 
\label{sub:in_an_ordered_phase}

In the bond-operator mean-field theory, an ordered state emerges when single $t_{\alpha}$ bosons condense on dimers at some $\vec{k}=\vec{k^\ast}$. Equivalently, this means that the existence of gapless excitation for triplons, i.e., $\omega_\vec{k^\ast}=0$. This vanishing spin-gap gives a condition $\Lambda_{\vec{k^\ast}}=\pm \Delta_{\vec{k^\ast}}$, which fixes the renormalized chemical potential $\mu$ as $  \mu_r=-2\left(\mathcal{A}\mp\mathcal{B}\right)\gamma_{\vec{k^\ast}}$. An obvious order parameter in an ordered phase would be the average number of condensed triplet bosons per dimer (or simply the triplon density), which we define: $ n_t \equiv \frac{1}{N} \ev{t_{\vec{k^\ast}\alpha}^\dagger t_{\vec{k^\ast}\alpha}}$ as in~\cite{Kumar2008}.

Taking summation over the position of dimers (i.e., $\vec{r}$) on both sides in the constraint equation~\eqref{eq:constraint}, and then performing Fourier transform, we get $\frac{1}{N} \sum_{\vec{k}}^{} t_{\vec{k}\alpha}^\dagger t_{\vec{k} \alpha}= 1-\bar{s}^2$. Further, we split the summation over $\vec{k}$ into two parts: $\sum_{\vec{k}}=\sum_{\vec{k}=\vec{k^\ast}}+\sum_{\vec{k}\neq \vec{k^\ast}}$. After taking the expectation value with respect to the ground state, the triplon density reduces to
\begin{align}
   n_t &= 1-\bar{s}^2 - \frac{3}{2N} \sum_{\vec{k}\neq \vec{k^\ast}}^{}\left(\frac{\Lambda_{\vec{k}}}{\omega_{\vec{k}}} -1\right),\label{eq:split_rt_exact}\\
   &\approx \frac{5}{2} -\bar{s}^2 - \frac{3}{2N} \sum_{\vec{k}\neq \vec{k^\ast}}^{} \frac{\Lambda_{\vec{k}}}{\omega_{\vec{k}}}.\label{eq:split_rt}
\end{align}
In Eq.~\eqref{eq:split_rt_exact}, we used the result $\ev{t_{\vec{k}\alpha}^\dagger t_{\vec{k} \alpha}}=v_{\vec{k}}^2=\left(\Lambda_{\vec{k}}/\omega_{\vec{k}}-1 \right)/2$, which one can get using canonical Bogoliubov transformation defined earlier. Performing the same splitting in the summation over $\vec{k}$ in the self-consistent equation~\eqref{eq:sce_ss}, we obtain
\begin{equation}
  \label{eq:split_ss}
  \frac{5}{2} -\bar{s}^2 - \frac{3}{2N} \sum_{\vec{k}\neq \vec{k^\ast}}^{} \frac{\Lambda_{\vec{k}}}{\omega_{\vec{k}}} = \frac{3}{2N}\frac{\Lambda_{\vec{k^\ast}}}{\omega_{\vec{k^\ast}}}.
\end{equation}
Using Eqs.~\eqref{eq:split_rt} and~\eqref{eq:split_ss}, we write the triplon density as
\begin{equation}
  \label{eq:rt_final}
   n_t \equiv \frac{1}{N} \ev{t_{\vec{k^\ast}\alpha}^\dagger t_{\vec{k^\ast}\alpha}} \approx \frac{3}{2N}\frac{\Lambda_{\vec{k^\ast}}}{\omega_{\vec{k^\ast}}}.
\end{equation}
It turns out that the gapless condition $\Lambda_{\vec{k^\ast}}=\Delta_{\vec{k^\ast}}$ does not produce a self-consistent equation for $n_t$ when one tries to derive it from Eq.~\eqref{eq:sce_mu} using~\eqref{eq:rt_final}. Therefore, we have not searched for a solution to this condition. The self-consistent equations with the condition $\Lambda_{\vec{k^\ast}}=-\Delta_{\vec{k^\ast}}$ (or, equivalently, $\mu_r=-2\left(\mathcal{A}+\mathcal{B}\right)\gamma_{\vec{k^\ast}}$) can thus be written as
\begin{align}
  n_t \xi &= J_D-\mu_r-\frac{3}{N}J_{-} \sum_{\vec{k}\neq \vec{k^\ast}}^{} \frac{\gamma_{\vec{k}}\left(\Lambda_{\vec{k}}-\Delta_{\vec{k}}\right)}{\omega_{\vec{k}}},\\
  \bar{s}^2 &= \frac{5}{2} - n_t - \frac{3}{2N} \sum_{\vec{k}\neq \vec{k^\ast}}^{} \frac{\Lambda_{\vec{k}}}{\omega_{\vec{k}}}, \\
 P&= n_t \gamma_{\vec{k^\ast}} +\frac{3}{2N} \sum_{\vec{k}\neq \vec{k^\ast}}^{} \frac{\gamma_{\vec{k}}\Lambda_{\vec{k}}}{\omega_{\vec{k}}}, \\
 Q&= n_t \gamma_{\vec{k^\ast}}-\frac{3}{2N} \sum_{\vec{k}\neq \vec{k^\ast}}^{} \frac{\gamma_{\vec{k}}\Delta_{\vec{k}}}{\omega_{\vec{k}}},
\end{align}
where $\xi=4J_{-}\gamma_{\vec{k^\ast}}$.



\section{Results and discussion} 
\label{sec:results_and_discussion}

This section presents the results of various physical quantities of interest in the dimer (paramagnetic) and ordered phases. We get these quantities by finding the numerical solutions of self-consistent equations in each phase in two categories: (i) in one case, we keep terms up to quadratic in triplet operators and ignore the higher order terms, and (ii) in the other case, we also include higher order terms (that is, quartic terms) using quadratic decouplings. In the latter case, the cubic terms do not contribute to the mean-field Hamiltonian due to the presence of the Levi-Civita tensor. Our primary interest is determining the boundaries of the ground-state phases that emerge due to phase transitions as one varies the coupling ratio $J_2/J_1$. The vanishing singlet-triplet spin-gap or triplon number density determines the boundary between a dimer phase and an ordered phase, with a continuous phase transition across the boundary. In our analyses, the lowest but nonzero value of energy dispersion $\omega_{\vec{k}^\ast}$ (finite singlet-triplet spin-gap) of the quasiparticles characterizes a dimer phase. In contrast, the staggered magnetization or triplon density specifies an antiferromagnetic ordered phase.

\begin{figure}[h]
  \centering
    \includegraphics[width=.45\textwidth]{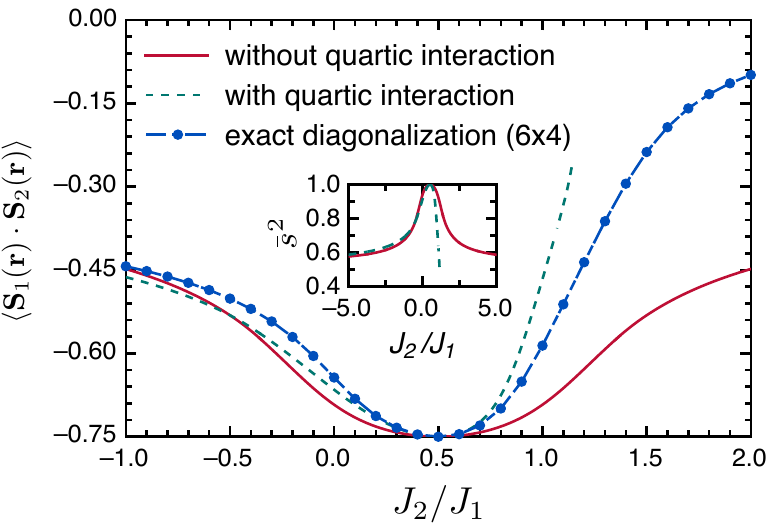}
  \caption{Intra-dimer spin-spin correlation, $\ev{\vec{S}_1(\vec{r})\cdot\vec{S}_2(\vec{r})} = \frac{1}{4}-\bar{s}^2$, from bond-operator mean-field theory. One gets this expression by taking ground state expectation on both sides of Eq.~\eqref{eq:sisj_intra} and then using the constraint ~\eqref{eq:constraint}. This figure also contains exact diagonalization data of averaged spin-spin correlations on dimers. The inset picture displays the mean-field data for the $\bar{s}^2$ parameter.}
  \label{fig:sisj}
\end{figure}

At the exact point, the square of the amplitude of singlet condensation on dimers acquires its maximum value of unity, or, equivalently, the intra-dimer spin-spin correlation gets the value $-3/4$ (Fig.~\ref{fig:sisj}) that corresponds to the eigenvalue of dimer operator $\vec{S}_1(\vec{r}) \cdot \vec{S}_2(\vec{r})$ corresponding to the spin-singlet state.

\begin{figure}[h]
  \centering
    \includegraphics[width=.235\textwidth]{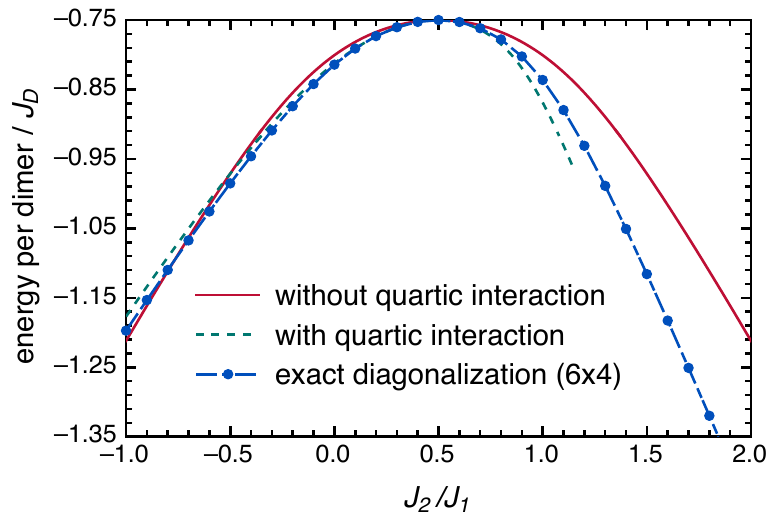}
    \includegraphics[width=.215\textwidth]{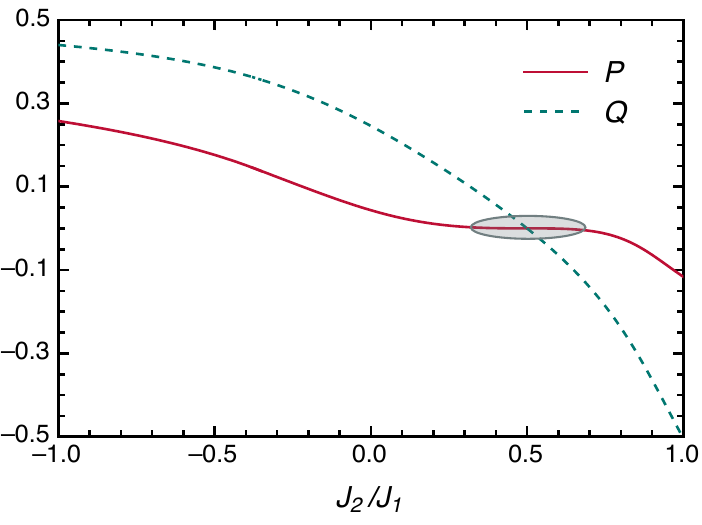}
  \caption{On the left, ground state energy per dimer, in units of $J_D$, from the bond-operator mean-field theory and exact diagonalization is shown in this picture, while on the right, the mean-field parameters $P$ and $Q$. In the shaded region of the right picture, the value of $P$ is almost zero.}
  \label{fig:EnergyPQ}
\end{figure}

Moreover, at the same coupling ratio, the ground state energy per dimer in units of $J_D$ yields the value $-3/4$ (Fig.~\ref{fig:EnergyPQ}). These mean-field results agree precisely with the numerical diagonalization results. This excellent agreement confirms the exactness of the dimer ground state $\ket{\Psi}$ at $J_2/J_1=1/2$. Furthermore, at this exchange ratio, the mean-field parameters $P$ and $Q$ vanish (see on the right in Fig.~\ref{fig:EnergyPQ}), and the chemical potential obtains the constant value $-3J_D/4$ (from Eq.~\eqref{eq:sce_mu}).

\begin{figure}[b]
  \centering
    \includegraphics[width=.38\textwidth]{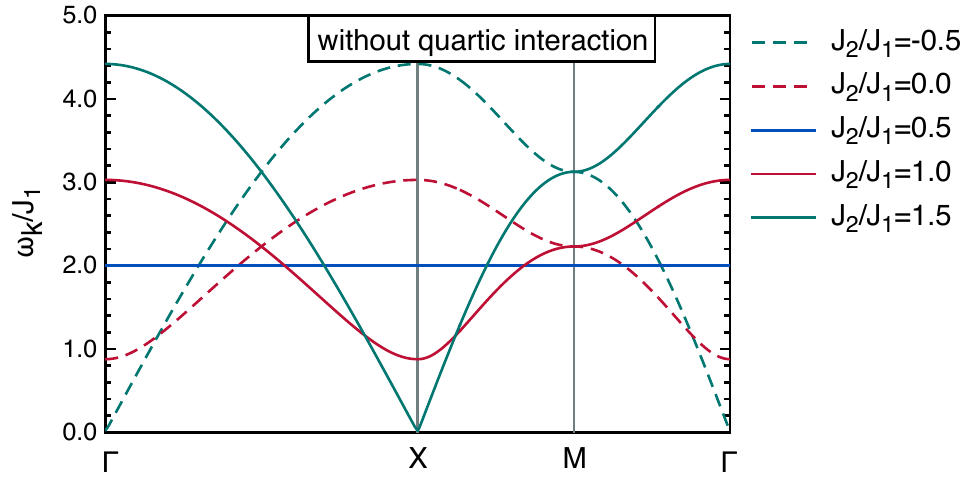}
    \includegraphics[width=.2\textwidth]{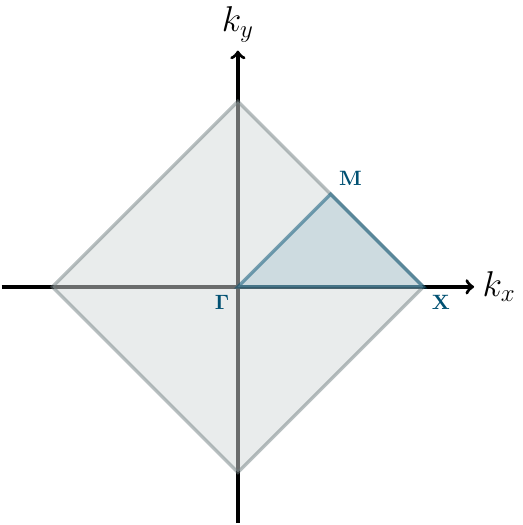}
    \includegraphics[width=.38\textwidth]{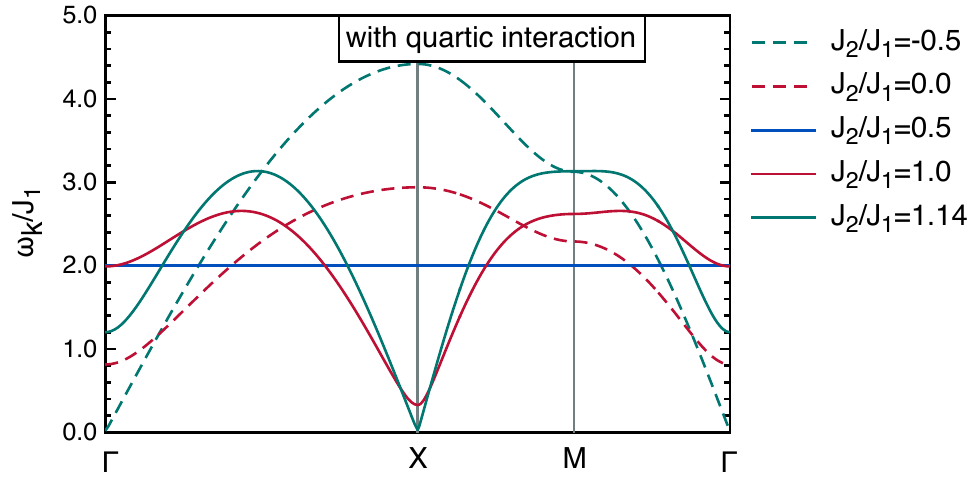}
  \caption{This picture exhibits the energy dispersion curves $\omega_{\vec{k}}$ (in units of $J_1$) with and without quartic interaction along the closed path connected by the highest symmetry points $\Gamma=(0,0)$, $X=(\pi,0)$, and $M=(\pi/2,\pi/2)$ of the magnetic Brillouin zone shown in the middle.}
  \label{fig:energy_dispersion}
\end{figure}

The triplet excitations at the exact point are gapped and localized, and these can be understood as follows. We get an immediate excited state $\ket{\Phi}$ when a single dimer of the product state $\ket{\Psi}$ excites from the spin-singlet state to a spin-triplet state. This process results in the creation of a single triplon quasi-particle. The energy cost to create a triplon, $\ev{H}{\Phi}-\ev{H}{\Psi}$, is just $J_D$ as the spin triads associated with the excited dimer are only affected. There are only four affected triads, each of which contributes $J_D/4$ to triplon excitation energy. At the exact point, a triplon stays localized on the excited dimer (see the flat curve in Fig.~\ref{fig:energy_dispersion}) as all the spin triads except the four associated with the excited dimer remain in the lowest spin state ($S_{tot}=1/2$, of a triad). The exact diagonalization results also exhibit that the first excitation state is a spin-$1$ state. However, the values of the singlet-triplet excitation gap differ quantitatively. In the diagonalization method, the gap comes close to 1.41 for a $6\times 4$ cluster with periodic boundary conditions. Generically, the bond-operator mean-field theory overestimates the spin-gap value in frustrated systems, as other works also found similar disagreements~\cite{Brenig1997, Bouzerar2001, Hwang2012, Doretto2014}.

\onecolumngrid 

\twocolumngrid

As we move away from the exact point, the singlet condensation amplitude on dimers falls off but always be non-zero at all coupling ratios (see the inset picture in Fig.~\ref{fig:sisj}). In harmonic approximation, the mean-field parameter $\bar{s}^2$ forms a bell-shaped symmetric curve about the exact point whose tails remain substantially off from zero even if we go far away from the exact point. We can understand this from the structural principles of the bond-operator mean-field theory, where we assume that there is always a background of singlet dimers and an ordered state emerges out of it when some dimers go into triplet states. In non-harmonic approximation, below the exact point, the square of singlet condensation amplitude remains close to the respective curve of the harmonic approximation. However, it sharply falls as one increases the ratio $J_2/J_1$  beyond the exact point. This unusual behavior originates because the mean-field parameter of particle-particle or hole-hole type  (that is, $Q$) decays much faster than the particle-hole typed parameter $P$ (see the right subfigure in Fig.~\ref{fig:EnergyPQ}). As a result, we get an asymmetric curve of $\bar{s}^2$ in the non-harmonic approximation. We also observe that the particle-hole type contribution stays almost negligible in a small vicinity about the exact point, highlighted by the shaded region shown on the right in Fig.~\ref{fig:EnergyPQ}.

\begin{figure}[h]
  \centering
    \includegraphics[width=.235\textwidth]{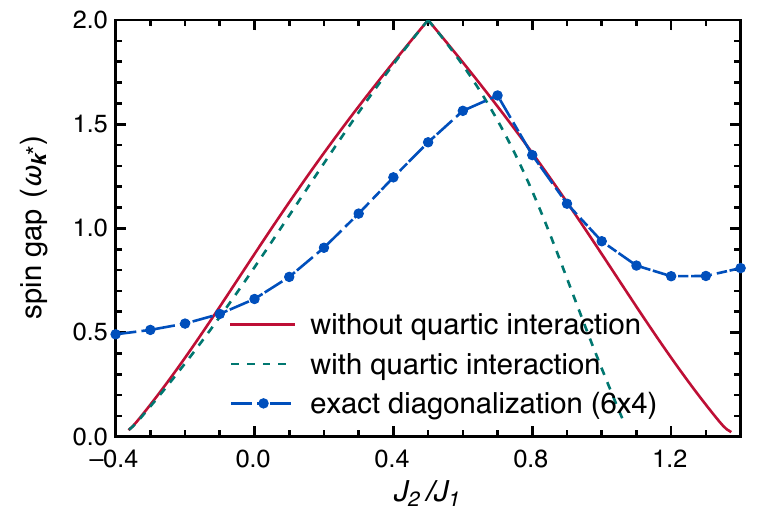}
    \includegraphics[width=.215\textwidth]{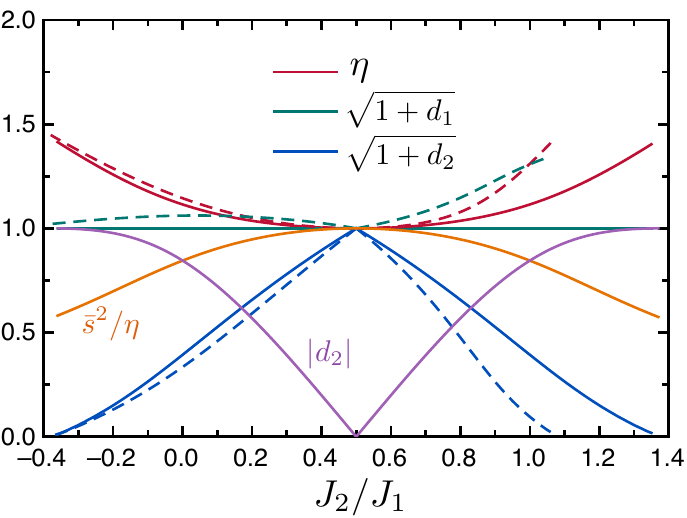}
  \caption{The left figure displays the singlet-triplet excitation gap calculated from bond-operator mean-field theory and exact diagonalization, while the right one shows plots for the variables involved in $\omega_k=J_D\eta\sqrt{(1+d_1 \gamma_\vec{k})(1+d_2 \gamma_\vec{k})}$. The solid and dashed lines in the right figure represent curves for harmonic and non-harmonic approximations, respectively.}
  \label{fig:spin_gap}
\end{figure}

In the dimer phase, the spin-gap is nonzero and decreases almost linearly as we move away from the exact point (Fig.~\ref{fig:spin_gap}). We can comprehend this linear drop by recasting the dispersion expression as $\omega_k=J_D\eta\sqrt{(1+d_1 \gamma_\vec{k})(1+d_2 \gamma_\vec{k})}$, where $\eta=1/4-\mu/J_D$, $d_1\eta=\rho_{+}(P+Q)$, and $d_2\eta=2\rho_{-}\bar{s}^2+\rho_{+}(P-Q)$ with $\rho_{\pm}=J_2/J_1\pm 1/2$. The $\vec{k^\ast}$ vectors that define the spin-gap (or minimum energy dispersion) are $\Gamma=(0,0)$ and $X=(\pi,0)$ for the dimer regions below and above the exact point, respectively (see energy dispersion curves in the Fig.~\ref{fig:energy_dispersion}). With these minimizing $k$-vectors, the spin gap takes the form $\Delta_{\pm}=J_D\eta\sqrt{(1\pm d_1)(1\pm d_2)}$, where plus and minus signs respectively correspond to $\Gamma$ and $X$. In the harmonic approximation, $d_1$ vanishes and $d_2=2\rho_{-}\bar{s}^2/\eta$, and, close to the exact point, the leading correction term in the expansion of $\sqrt{1\pm d_2}$ is linear in $J_2/J_1$ as $\bar{s}^2\sim 1$ and also $\eta\sim1$ (see Eq.\eqref{eq:sce_mu}). As one goes away from the exact point, the $|d_2|$ initially increases linearly with $J_2/J_1$ and then saturates to unity. The decrement in $\bar{s}^2$ is more rapid than the increment in the chemical potential. As a result, the ratio $\bar{s}^2/\eta$ falls off on either side of the exact point (see Fig.~\ref{fig:spin_gap}). This quadratic fall compensates for the linear gain in $|\rho_{-}|$, so the $|d_2|$ remains less than unity in the dimer region. The higher order correction terms in the series of $\sqrt{1\pm d_2}$, therefore, do not contribute significantly in the entire dimer phase, and so the $\sqrt{1\pm d_2}$ decays linearly with $J_2/J_1$ in the gapped phase. Moreover, the expansion of $\eta$ has leading correction term quadratic in $J_2/J_1$, that is, $\eta=1+\mathcal{O}(J_2/J_1)^2$ (see Fig.~\ref{fig:spin_gap}). Thus, the product of the series of $\eta$ and $\sqrt{1\pm d_2}$ gives nearly a linear decline in the spin gap. Using Eq.~\eqref{eq:sce_mu}, one may find an analytical expansion of $\eta$ in terms of $J_2/J_1$, and we expect a rigor calculation here that involves the complete elliptic integrals as done in the case of spin ladders~\cite{Gopalan1994}. Moreover, we can also understand linearity in the spin gap for the non-harmonic case, although more intricacies are involved here due to the mean-field parameters $P$ and $Q$. One notices that the spin-gap curve is symmetric about the exact point in harmonic approximation as in singlet condensation amplitude and ground state energy. In contrast, the spin gap curve is asymmetric in non-harmonic approximation due to a steep fall in $Q$ above the exact point (see Fig.~\ref{fig:EnergyPQ}).

\begin{figure}[h]
  \centering
    \includegraphics[width=.24\textwidth]{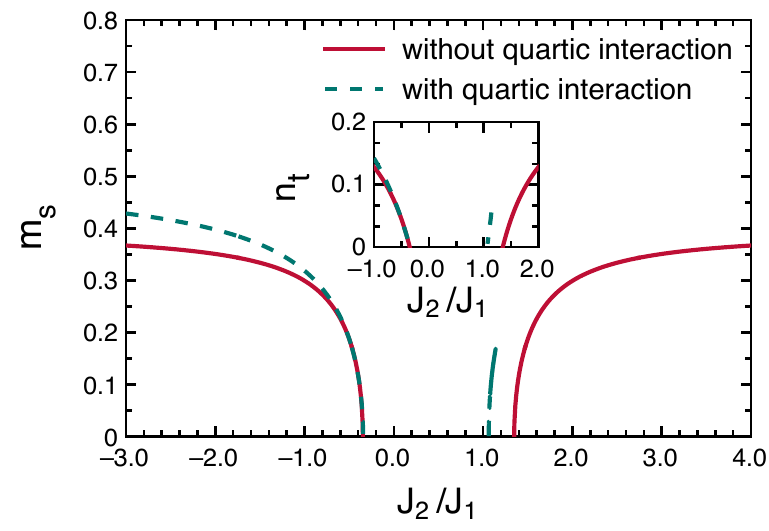}
    \includegraphics[width=.213\textwidth]{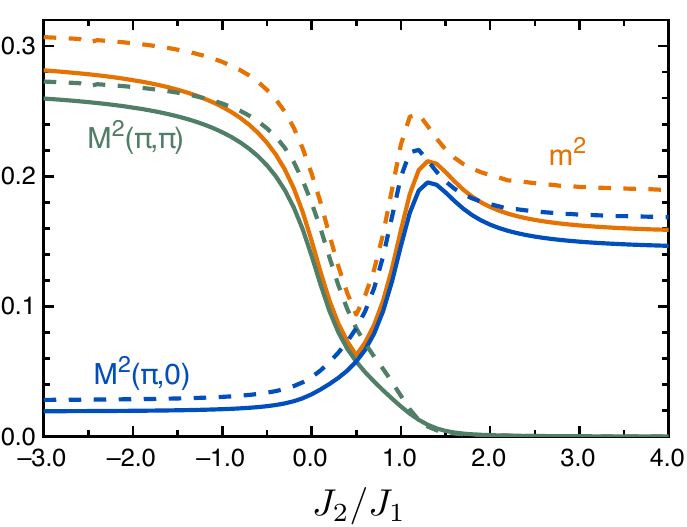}
  \caption{Left: the staggered magnetization and triplon number density found from bond-operator mean-field theory are shown. Right: The order parameters $M^2(\vec{k})$ and $m^2$ from exact diagonalization are displayed; here, the solid and dashed lines, respectively, represent systems $6 \times 4$ and $4 \times 4$ with periodic boundary conditions.}
  \label{fig:triplon_density}
\end{figure}

\begin{figure}[h]
  \centering
    \includegraphics[width=.45\textwidth]{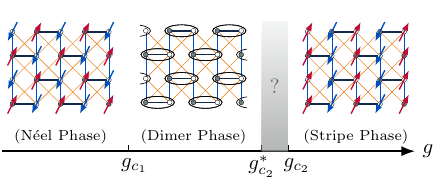}
  \caption{Quantum phase diagram. Filled and empty small circles denote the sub-lattices of a square lattice. In our mean-field calculation, we label $1$ to filled circles and $2$ to empty ones. The values of critical points are: $g_{c_1}\approx-0.35$, $g_{c_2}\approx 1.35$, and $g_{c_2}^\ast\approx 1.07$.}
  \label{fig:PhaseDiagram}
\end{figure}

In the ordered phases, we calculate and analyze different order parameters designed to detect the robustness of an antiferromagnetic state. The spin gap data and dispersion curves calculated from the bond-operator mean-field theory indicate two antiferromagnetic orders with $\Gamma$ and $X$ wave vectors. These vectors are associated with the staggered dimer lattice (Fig.~\ref{fig:dimer_lattices}) and the sub-lattice labeling. We use triplon number density and staggered magnetization ($m_s=\bar{s}\sqrt{n_t}$, see Ref.~\cite{Kumar2008} for derivation) to measure the strength of antiferromagnetic orders. These order parameters are shown on the left in Fig.~\ref{fig:triplon_density}. The wave vectors $\Gamma$ and $X$ correspond to the $(\pi,\pi)$ and $(\pi,0)$ ordering wave vectors for a square lattice. The N\'eel and stripe phases shown in Fig.~\ref{fig:PhaseDiagram} are associated with the ordering vectors. In the harmonic approximation, the N\'eel phase emerges as $J_2/J_1$ goes below $-0.35$, and the stripe (or collinear) phase starts developing from $J_2/J_1=1.35$. However, the non-harmonic approximation extends the domain of the stripe phase in the phase diagram. Recently, a large-scale quantum Monte Carlo study of a spatially anisotropic Heisenberg antiferromagnet on the honeycomb lattice has been done~\cite{Sushchyev2023}. This research suggests that our model should be in the staggered dimer phase if one turns off the next-neighbor coupling $J_2$. Our bond-operator mean-field results indeed agree with this prediction.

We also employ numerical exact diagonalization on finite clusters to inspect the legitimacy of the mean-field outcomes. There are two types of antiferromagnetic order parameters available. In one construction, one defines $m^2=\frac{1}{\mathcal{N}^2} \sum_{i,j}^{}|\ev{\vec{S}_i\cdot\vec{S}_j}|$, where $\mathcal{N}$ is the total number of lattice sites (see Ref.~\cite{Richter2004}). It measures the firmness or weakness of an antiferromagnetic order but does not identify the ordering wave vector. In other design, we define $\vec{k}$-dependent magnetic susceptibility as $M^2(\vec{k})=\frac{1}{\mathcal{N}(\mathcal{N}+2)} \sum_{i,j}^{} \ev{\vec{S}_i\cdot\vec{S}_j}e^{i\vec{k}\cdot (\vec{x}_i-\vec{x}_j)}$, where $\vec{x}_i$ is the position of $i^{th}$ spin (see Ref.~\cite{Schulz1996}). We show these order parameters for N\'eel and stripe phases on the right in Fig.~\ref{fig:triplon_density}. The $M^2(\pi,\pi)$ and $M^2(\pi,0)$ are robust in N\'eel and stripe phases, respectively. At the same time, $m^2$ gets high values in both the phases. We also calculate these order parameters with open boundary conditions along the vertical direction for the $4\times 7$ system. Again, N\'eel and stripe phases prevail as one goes far away from the exact point. Moreover, the exact diagonalization data suggest that the exact point is paramagnetic in the thermodynamic limit.


\section{Conclusions} 
\label{sec:conclusions}

In this work, we introduced a spatially anisotropic Heisenberg magnet with bilinear nearest and next-nearest neighbor interactions, which has an exact staggered dimer ground state at $J_2/J_1=1/2$. Here, the nearest exchange interactions are antiferromagnetic, while the next-nearest exchange interactions may be either ferromagnetic or antiferromagnetic. We performed the bond-operator mean-field formulation and numerical diagonalization of the dimerized magnet. At the exact point, these analyses agree with the exact results. According to the mean-field theory, the magnet goes into magnetically ordered phases (N\'eel and Stripe) through continuous phase transitions as one moves away on either side of the exact point. Mean-field theory, including the higher-order triplet terms, slightly modifies the phase boundary for the stripe phase. Exact diagonalization results also agree with the nature of dimer and ordered phases.

\begin{figure}[b]
  \centering
    \includegraphics[width=.9\textwidth]{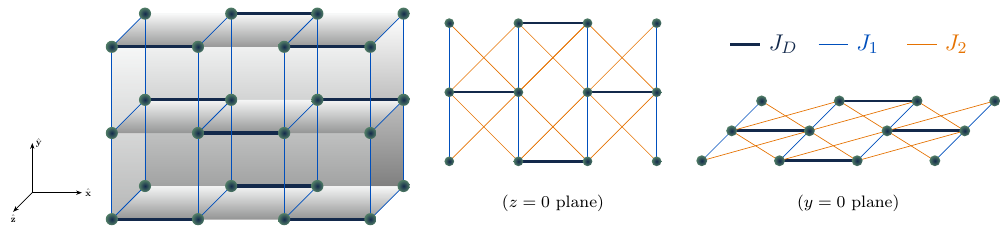}
  \caption{Second picture from the left represents a three-dimensional extension of the model~\eqref{eq:model_hamiltonian}, where the next neighbor $J_2$ bonds are not drawn for clarity. The last two pictures represent lattice layers in $xy$- and $zx$-planes. The $yz$-plane is not shown explicitly, as it is without dimers. The first picture fixes the coordinate axes.}
  \label{fig:Model3D}
\end{figure}

The above model can also be extended in three dimensions but with a different solvable point for an exact staggered dimer ground state. We show such a possibility on the left in Fig.~\ref{fig:Model3D}. Here, the two-dimensional staggered dimer lattices of the type shown in Fig.~\ref{fig:dimer_lattices} are stacked directly above one another along the $z$-axis such that all even layers (imagine \emph{infinite} in extent) are translated by the same $a\hat{y}$ amount. The lattice planes normal to the $y$- and $z$-axes have dimers. However, the lattice planes perpendicular to the $x$-axis are without dimers and form an isotropic Heisenberg model with \emph{only} nearest-neighbor interactions on square lattices. We expect that this proposed model on the three-dimensional lattice with exchange interactions $J_D$, $J_1$, $J_2$ will have a staggered dimer ground state when one sets $J_D=4J_1$ and $J_2/J_1=1/2$. It is because the number of triangles that share the same dimer becomes double the corresponding value in the two-dimensional model~\ref{eq:model_hamiltonian}, while nothing changes for non-dimer bonds. 

\onecolumngrid 

\twocolumngrid

In the last many years, the exact dimer models have renewed interest for researchers due to theoretical and technological advancements. In that regard, the proposed bilinear models can be helpful in many ways. One immediate investigation, though more accurate and reliable tools, could be to examine the nature of phase transition. Whether the transitions fall in the Landau-Ginzburg-Wilson paradigm or are of the deconfined quantum criticality class, future probes can thoroughly address such fundamental questions. Lastly, we emphasize that several similar models can also be designed on the bigger unit cells. We will address a few in separate papers.


\section*{Acknowledgements} 
\label{sec:acknowledgements}

R.K. sincerely thanks Brijesh Kumar (JNU, India) for the valuable suggestions and for pointing out their relevant work. Manas Ranjan Mahapatra acknowledges the financial support from the Central University of Rajasthan, Ajmer (India).

\bibliography{stagdim}

\end{document}